\documentclass[conference,twocolumn]{IEEEtran}
\usepackage{multicol}
\usepackage{amsmath}
\usepackage{graphicx}
\usepackage{amsfonts}
\usepackage{epsfig}
\usepackage{epstopdf}
\usepackage{psfrag}
\begin{document}

\title{Transmission Control of Two-User Slotted ALOHA Over Gilbert-Elliott Channel: Stability and Delay Analysis}

\author{\IEEEauthorblockN{Anthony Fanous\
and Anthony Ephremides}
\IEEEauthorblockA{Department of Electrical and Computer Engineering\
\\University of Maryland, College Park, MD 20742}
\{afanous, etony\}@umd.edu}

\maketitle
\begin{abstract}
\textbf{In this paper, we consider the problem of calculating the stability region and average delay of two user slotted ALOHA over a Gilbert-Elliott channel, where users have channel state information and adapt their transmission probabilities according to the channel state. Each channel has two states, namely, the 'good' and 'bad' states. In the 'bad' state, the channel is assumed to be in deep fade and the transmission fails with probability one, while in the 'good' state, there is some positive success probability. We calculate the stability region with and without Multipacket Reception capability as well as the average delay without MPR. Our results show that the stability region of the controlled S-ALOHA is always a superset of the stability region of uncontrolled S-ALOHA. Moreover, if the channel tends to be in the 'bad' state for long proportion of time, then the stability region is a convex polygon strictly containing the TDMA stability region and the optimal transmission strategy is to transmit with probability one whenever the nodes have packets and it is shown that this strategy is delay optimal. On the other hand, if the channel tends to be in the 'good' state more often, then the boundary of the stability region is characterized by a convex curve and is strict subset of the TDMA stability region. We also show that enhancing the physical layer by allowing MPR capability can significantly enhance the performance while simplifying the MAC Layer design by the lack of the need of scheduling under some conditions. Furthermore, it is shown that transmission control not only allows handling higher stable arrival rates but also leads to lower delay for the same arrival rate compared with ordinary S-ALOHA.}
\end{abstract}

\IEEEpeerreviewmaketitle

\section{Introduction}
Random Access is preferred in large wireless networks as it does not need any coordination between the nodes, which largely simplifies the MAC-Layer Protocol Design. However, Random Access schemes - despite their simplicity - were known to be suboptimal compared to orthogonal access schemes such as TDMA over the collision channel. ALOHA was first initiated by the work of Abramson [1]. Tsybakov and Mikhailov derived sufficient conditions on the stability of two user ALOHA by using the idea of stochastic dominance [2]. Rao and Ephremides in [3] used the idea of dominant systems to decouple the interaction between the queues and derive the exact stability region of two user S-ALOHA over collision channel as well as inner bounds for the $N>$2 case. Later, Luo and Ephremides [4] introduced the idea of stability ranks to derive tight bounds on the stability region over collision channel for $N>$2 case. Ghez, Verdu and Schwartz in [5] analysed ALOHA with MPR capability under infinite user and single buffer model where users are indistinguishable. Stability region of S-ALOHA with MPR capability in a non-symmetric configuration was first derived by Naware, Mergen and Tong [6] in which they showed that by improving the MPR capability, the stability region undergoes a phase transition from the concave region to a convex polyhedron and in this case, S-ALOHA outperforms TDMA and was shown to be optimal. Adireddy and Tong [7] considered the effect of knowledge of channel state information (CSI) in an $N$ user symmetric S-ALOHA on the maximum aggregate stable throughput rate. In this paper, we consider an asymmetric S-ALOHA system with time varying links according to Gilbert-Elliott model [8]. Users have exact channel knowledge and adjust their transmission probabilities according to the channel state. We calculate the stability region with and without MPR capability as well as the average delay without MPR. The main result is that S-ALOHA with transmission control -from a stability or delay point of view- outperforms TDMA whenever the channels tend to be in the bad state and in this case there is no need for scheduling as the optimal strategy is to transmit whenever backlogged. Moreover, by enhancing the physical layer by allowing MPR, S-ALOHA with transmission control can outperform TDMA even if the channel does not have tendency to be in the bad state for long proportion of time which attracts the attention to the capability of random access with transmission control over time varying channels which makes it suitable to use over networks that lack strong coordination between the users.\\
The paper is organized as follows: In section II, we introduce the channel model. In section III, we calculate the stability region of a controlled two user S-ALOHA without MPR. In section IV, we consider the effect of Multipacket reception capability (MPR) on the stability region. In section V, we consider the minimum average delay per packet without MPR and in section VI we conclude the paper.
\section{System Model}
The system consists of an uplink with two source nodes and one destination node. Time is slotted with slot duration equals to one packet duration. Arrivals to user $i$ occur according to a Bernoulli process with parameter $\lambda_i$, $i=1,2$ and are assumed to be independent between users and over slots. Each user has an infinite buffer for storing his packets. Channel is assumed to be independent between users and to vary between slots according to a Gilbert-Elliott model, where it can be in one of two states at any given time slot: the 'good' state that we denote by '1' and the bad state that we denote by '0'. Channel is assumed to be in the same state during a slot duration. Channel parameters are different between the users to account for the case when - for instance - one user is closer to the destination than the other so his channel remains in the 'good' state for a longer portion of time. The long term proportion of time in which user's $j$ channel is in state $i$ is denoted by $\pi_i^{(j)}$, $i=0,1$, $j=1,2$ and can be directly obtained by solving for the stationary probabilities of the Markov Chain describing the channel. We also define the transmission probabilities as function of the channel state as follows: $q_{ij}$ denotes the probability that user $j$ transmits given that his channel is in state $i$. We will denote by $f_{ij}$ the success probability when the channel of user $j$ is in state $i$. In this paper, we specialize to the case where the channel in 'bad' state is in deep fade and transmission is assumed to fail with probability one, i.e. $f_{0j}=0$, $j=1,2$. This assumption is for example in conform with the SNR threshold model for reception in which a packet is successfully decoded at the destination if and only if the SNR exceeds some threshold value. In the 'bad' state, the SNR is assumed to be below the threshold and hence the success probability is zero; while in the 'good' state SNR is above the reception threshold and hence the reception is successful with probability one. We relax the latter assumption by allowing some positive success probability whenever the channel is in the 'good' state.

\section{Stability Region and Delay without MPR}
In this section, we consider the case where the destination uses a simple receiver that does not have any MPR capability so that if both users transmit together, a collision occurs and neither of the packets can be successfully received. In order to calculate the stability region, we will use the notion of dominant systems as in [3], [4] and [6] to decouple the
interaction between the users' queues.\\
\begin{figure*}[!t]
\centering
\includegraphics[bb=0 0 1474 451,width=1\textwidth]{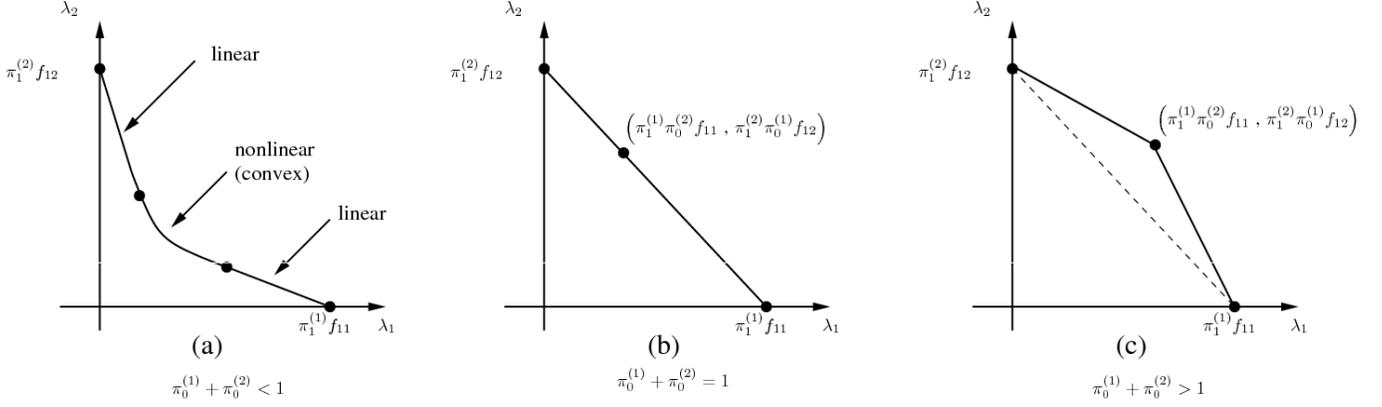}
\caption{Stability Region without MPR for various values of stationary probabilities}
\end{figure*}
We denote by $S_1$ the first dominant system. In $S_1$, arrivals to the queues as well as channel variations are assumed to be identical to those in the original system. However, in $S_1$, whenever user $1$'s queue empties, he will continue transmitting \textit{dummy} packets and hence causing more collisions with user $2$'s packets. First we note that in $S_1$, queues are no shorter than in the original system and hence stability region of the dominant system is a subset of the stability region of the original system.  Also, by using an argument of indistinguishability at saturation as in [3], we can conclude that the stability region of $S_1$ is a superset of the stability region of the original system, and hence both regions coincide for fixed transmission probabilities. It is clear that in $S_1$, $Q_1$ never empties and hence $Q_2$ sees a constant service rate while $Q_1$ service rate depends on the state of $Q_2$: empty or not. Specifically, the first dominant system is formulated as:
 \begin{multline}
 \lambda_2 < \mu_2=\mathrm{\mbox{Pr[User 2 is successful in a slot]}}\\
= \left(\pi_1^{(2)}q_{12}f_{12}\right)\left(1-\pi_0^{(1)}q_{01}-\pi_1^{(1)}q_{11}\right)
 \end{multline}
 On the other hand, $Q_1$ service rate will depend on the state of the other queue, specifically:
 \begin{multline}
 \lambda_1 < \left(1-\frac{\lambda_2}{\mu_2}\right)\left(\pi_1^{(1)}q_{11}f_{11}\right)\\
 +\left(\frac{\lambda_2}{\mu_2}\right)
\left(\pi_1^{(1)}q_{11}f_{11}\right)\left(1-\pi_0^{(2)}q_{02}-\pi_1^{(2)}q_{12}\right)
 \end{multline}
 where $\left(\frac{\lambda_2}{\mu_2}\right)$ and $\left(1-\frac{\lambda_2}{\mu_2}\right)$ are the probability that $Q_2$ is busy or idle in a slot respectively.\\
 Equivalently, $S_1$ can be written as:
 \begin{equation}
 \lambda_2 < \left(\pi_1^{(2)}q_{12}f_{12}\right)\left(1-\pi_0^{(1)}q_{01}-\pi_1^{(1)}q_{11}\right)
 \end{equation}
 \begin{equation}
 \lambda_1 < \left(\pi_1^{(1)}q_{11}f_{11}\right)\left[1-\left(\frac{\lambda_2}{\mu_2}\right)\left(\pi_0^{(2)}q_{02}+\pi_1^{(2)}q_{12}\right)\right]
\end{equation}
Similarly for $S_2$ in which $Q_2$ transmits dummy packets:
\begin{equation}
\lambda_1 < \mu_1=\left(\pi_1^{(1)}q_{11}f_{11}\right)\left(1-\pi_0^{(2)}q_{02}-\pi_1^{(2)}q_{12}\right)
\end{equation}
\begin{equation}
 \lambda_2 < \left(\pi_1^{(2)}q_{12}f_{12}\right)\left[1-\left(\frac{\lambda_1}{\mu_1}\right)\left(\pi_0^{(1)}q_{01}+\pi_1^{(1)}q_{11}\right)\right]
\end{equation}
It can be easily shown that for optimality, $q_{01}^*=q_{02}^*=0$  as they lead to a strictly higher stability region as we intuitively expect because transmission whenever the channel is in the bad state is unsuccessful with probability one.\\
The stability region is then given by:
\begin{equation}
\mathcal{S}=\bigcup_{(q_{11},q_{12})\in[0,1]^2}\mathcal{S}\mathrm{\mathit{(q_{11},q_{12})}}
\end{equation}
Where $\mathcal{S}\mathrm{\mathit{(q_{11},q_{12})}}$ is the stability region for fixed transmission probabilities $q_{11}$ and $q_{12}$ and is given by equations (3), (4), (5) and (6).
The problem of calculating the boundary of the stability region can be formulated as a constrained optimization problem which can be directly solved by using the same technique as in [9]. The detailed solution is presented in Appendix A.\\
\textit{Lemma 1:}\\
-\underline{If $\pi_0^{(1)}+\pi_0^{(2)} < 1$ :}\\
The boundary of the stability region is characterized by straight lines near the axes, and by strictly convex function in the middle part. The resulting stability region is given by:\\
\begin{equation}
\mathcal{R}=\mathcal{L}_1\bigcup\mathcal{L}_2\bigcup\mathcal{L}_3
\end{equation}
Where:
\begin{multline}
\mathcal{L}_1=\left\{(\lambda_1,\lambda_2):\lambda_2<\pi_1^{(2)}f_{12}-\left(\frac{\pi_1^{(2)}f_{12}}{\pi_0^{(2)}f_{11}}\right)\lambda_1,\right.\\
\mbox{  for  } \lambda_1 \in \left[0,(\pi_0^{(2)})^2f_{11}\right)\Bigg\}
\end{multline}
\begin{multline}
\mathcal{L}_2=\left\{(\lambda_1,\lambda_2):\sqrt{\frac{\lambda_1}{f_{11}}}+\sqrt{\frac{\lambda_2}{f_{12}}}<1,\right.\\
\mbox{  for  } \lambda_1 \in \left[(\pi_0^{(2)})^2f_{11},(\pi_1^{(1)})^2f_{11}\right)\Bigg\}
\end{multline}
\begin{multline}
\mathcal{L}_3=\left\{(\lambda_1,\lambda_2):\lambda_2<\pi_0^{(1)}f_{12}-\left(\frac{\pi_0^{(1)}f_{12}}{\pi_1^{(1)}f_{11}}\right)\lambda_1,\right.\\
\mbox{  for  } \lambda_1 \in \left[(\pi_1^{(1)})^2f_{11},\pi_1^{(1)}f_{11}\right)\Bigg\}
\end{multline}
-\underline{If $\pi_0^{(1)}+\pi_0^{(2)} \geq 1$ :}\\
The stability region is a convex polygon whose boundary is determined by two lines. The optimal transmission probabilities in this case are $(q_{11}^*,q_{12}^*)=(1,1)$. The resulting stability region is convex and given by:\\
\begin{equation}
\mathcal{R}=\mathcal{L}_1\bigcup\mathcal{L}_2
\end{equation}
Where:
\begin{multline}
\mathcal{L}_1=\left\{(\lambda_1,\lambda_2):\lambda_2<\pi_1^{(2)}f_{12}-\left(\frac{\pi_1^{(2)}f_{12}}{\pi_0^{(2)}f_{11}}\right)\lambda_1,\right.\\
\mbox{  for  } \lambda_1 \in \left[0,\pi_1^{(1)}(1-\pi_1^{(2)})f_{11}\right)\Bigg\}
\end{multline}
\begin{multline}
\mathcal{L}_2=\left\{(\lambda_1,\lambda_2):\lambda_2<\pi_0^{(1)}f_{12}-\left(\frac{\pi_0^{(1)}f_{12}}{\pi_1^{(1)}f_{11}}\right)\lambda_1,\right.\\
\mbox{  for  } \lambda_1 \in \left[\pi_1^{(1)}(1-\pi_1^{(2)})f_{11},\pi_1^{(1)}f_{11}\right)\Bigg\}
\end{multline}
\textit{Proof:} See Appendix A.\\
From Figure 1, we notice that whenever $\pi_0^{(1)}+\pi_0^{(2)}<1$ which roughly means that the channels tend to be in the good state, the stability region is strict subset of TDMA stability region but is strict superset of the stability region of ordinary S-ALOHA without transmission control given by $\sqrt{\frac{\lambda_1}{\pi_1^{(1)}f_{11}}}+\sqrt{\frac{\lambda_2}{\pi_1^{(2)}f_{12}}}=1$. If  $\pi_0^{(1)}+\pi_0^{(2)}=1$, the stability region becomes linear and coincides with the TDMA stability region. Finally, whenever $\pi_0^{(1)}+\pi_0^{(2)}>1$, the stability region becomes a convex region strictly containing the TDMA stability region, meaning that whenever the channel has tendency to be in the bad state, random access with transmission control outperforms orthogonal access.
\section{Effect of MPR Capability}
The Multipacket reception capability has a significant effect on the stability region of S-ALOHA with transmission control. Depending on the strength of the MPR, the effect can be either in a strict increase of the stability region without a phase transition or it can be in an increase of the stability region with phase transition. For example, without MPR, if $\pi_0^{(1)}+\pi_0^{(2)}<1$, TDMA outperforms S-ALOHA as we saw in last section. However, S-ALOHA with MPR in this case can also outperform TDMA.\\
Define $\tilde{f_i}$ to be the probability of success of the $i$th user whenever both users transmit simultaneously, which is typically zero if the receiver does not have any MPR capability. Simultaneous success of packets occur only if both channels are in the good state as transmission of a user fails with probability one if his channel is in the bad state.\\
By using the dominant system approach and again using that for optimality $q_{01}^*=q_{02}^*=0$, we can get the first dominant system in which $Q_1$ transmits $dummy$ packets as:\\
\begin{equation}
\lambda_2 < \left(\pi_1^{(2)}q_{12}f_{12}\right)\left[1-\pi_1^{(1)}q_{11}\left(1-\frac{\tilde{f_2}}{f_{12}}\right)\right]
\end{equation}
\begin{equation}
\lambda_1 < \left(\pi_1^{(1)}q_{11}f_{11}\right)\left[1-\frac{\left(1-\frac{\tilde{f_1}}{f_{11}}\right)\lambda_2}{f_{12}\left[1-\pi_1^{(1)}q_{11}\left(1-\frac{\tilde{f_2}}{f_{12}}\right)\right]}\right]
\end{equation}
Similarly for $S_2$ in which $Q_2$ transmits dummy packets:
\begin{equation}
\lambda_1 < \left(\pi_1^{(1)}q_{11}f_{11}\right)\left[1-\pi_1^{(2)}q_{12}\left(1-\frac{\tilde{f_1}}{f_{11}}\right)\right]
\end{equation}
\begin{equation}
\lambda_2 < \left(\pi_1^{(2)}q_{12}f_{12}\right)\left[1-\frac{\left(1-\frac{\tilde{f_2}}{f_{12}}\right)\lambda_1}{f_{11}\left[1-\pi_1^{(2)}q_{12}\left(1-\frac{\tilde{f_1}}{f_{11}}\right)\right]}\right]
\end{equation}
By following similar steps as in Appendix A, we can calculate the stability region of S-ALOHA with transmission control and MPR capability as:\\
\textit{Lemma 2:}\\
-\underline{If $\pi_0^{(1)}+\pi_0^{(2)}+\pi_1^{(2)}\frac{\tilde{f_1}}{f_{11}}+\pi_1^{(1)}\frac{\tilde{f_2}}{f_{12}} < 1$ :}\\
The boundary of the stability region is characterized by straight lines near the axes, and by strictly convex function in the middle part. The resulting stability region is given by:\\
\begin{equation}
\mathcal{R}=\mathcal{L}_1\bigcup\mathcal{L}_2\bigcup\mathcal{L}_3
\end{equation}
Where:
\begin{multline}
\mathcal{L}_1=\left\{(\lambda_1,\lambda_2):\frac{\lambda_2}{\pi_1^{(2)}f_{12}}+\frac{\left(1-\frac{\tilde{f_2}}{f_{12}}\right)\lambda_1}{f_{11}\left[1-\pi_1^{(2)}\left(1-\frac{\tilde{f_1}}{f_{11}}\right)\right]}<1,\right.\\
\left.\mbox{  for  } \lambda_1 \in \left[0,\frac{f_{11}\left[1-\pi_1^{(2)}\left(1-\frac{\tilde{f_1}}{f_{11}}\right)\right]^2}{\left(1-\frac{\tilde{f_2}}{f_{12}}\right)}\right)\right\}
\end{multline}
\begin{multline}
\mathcal{L}_2=\left\{(\lambda_1,\lambda_2):\sqrt{\frac{\left(1-\frac{\tilde{f_2}}{f_{12}}\right)\lambda_1}{f_{11}}}+\sqrt{\frac{\left(1-\frac{\tilde{f_1}}{f_{11}}\right)\lambda_2}{f_{12}}}<1,\right.\\
\left.\mbox{  for  } \lambda_1 \!\in \! \!\left[\!\frac{f_{11}\!\left[1\!-\!\pi_1^{(2)}\!\!\left(1-\frac{\tilde{f_1}}{f_{11}}\right)\right]^2}{\left(1-\frac{\tilde{f_2}}{f_{12}}\right)},(\pi_1^{(1)})^2f_{11}\!\!\left(\!\!1\!-\!\frac{\tilde{f_2}}{f_{12}}\right)\!\right)\!\right\}
\end{multline}
\begin{multline}
\mathcal{L}_3=\left\{(\lambda_1,\lambda_2):\frac{\lambda_1}{\pi_1^{(1)}f_{11}}+\frac{\left(1-\frac{\tilde{f_1}}{f_{11}}\right)\lambda_2}{f_{12}\left[1-\pi_1^{(1)}\left(1-\frac{\tilde{f_2}}{f_{12}}\right)\right]}<1,\right.\\
\left.\mbox{  for  } \lambda_1 \in \left[(\pi_1^{(1)})^2f_{11}\left(1-\frac{\tilde{f_2}}{f_{12}}\right),\pi_1^{(1)}f_{11}\right)\right\}
\end{multline}
-\underline{If $\pi_0^{(1)}+\pi_0^{(2)}+\pi_1^{(2)}\frac{\tilde{f_1}}{f_{11}}+\pi_1^{(1)}\frac{\tilde{f_2}}{f_{12}} \geq 1$ :}\\
The stability region is a convex polygon whose boundary is determined by two lines. The optimal transmission probabilities in this case are $(q_{11}^*,q_{12}^*)=(1,1)$. The resulting stability region is convex and given by:\\
\begin{equation} \mathcal{R}=\mathcal{L}_1\bigcup\mathcal{L}_2
\end{equation}
Where:
\begin{multline}
\mathcal{L}_1=\left\{(\lambda_1,\lambda_2):\frac{\lambda_2}{\pi_1^{(2)}f_{12}}+\frac{\left(1-\frac{\tilde{f_2}}{f_{12}}\right)\lambda_1}{f_{11}\left[1-\pi_1^{(2)}\left(1-\frac{\tilde{f_1}}{f_{11}}\right)\right]}<1,\right.\\
\left.\mbox{  for  } \lambda_1 \in \left[0,\pi_1^{(1)}f_{11}\left[1-\pi_1^{(2)}\left(1-\frac{\tilde{f_1}}{f_{11}}\right)\right]\right)\right\}
\end{multline}
\begin{multline}
\mathcal{L}_2=\left\{(\lambda_1,\lambda_2):\frac{\lambda_1}{\pi_1^{(1)}f_{11}}+\frac{\left(1-\frac{\tilde{f_1}}{f_{11}}\right)\lambda_2}{f_{12}\left[1-\pi_1^{(1)}\left(1-\frac{\tilde{f_2}}{f_{12}}\right)\right]}<1,\right.\\
\left.\mbox{  for  } \lambda_1 \!\in\! \left[\pi_1^{(1)}\!f_{11}\!\left[1\!-\!\pi_1^{(2)}\!\left(1\!-\!\frac{\tilde{f_1}}{f_{11}}\right)\right]\!,\pi_1^{(1)}f_{11}\right)\right\}
\end{multline}
This raises the attention that by enhancing the physical layer capabilities of the receivers by allowing MPR capability, random access can outperform TDMA over time varying channels even though it needs little or no coordination between the users.
\section{Delay Analysis}
In this section, we consider the delay analysis of a symmetric two-user S-ALOHA system with transmission control over Gilbert-Elliott channel under the simplified reception model stated before. By $symmetry$ we mean that the average arrival rates to both users are identical and both users' channel conditions are identical and hence the users are indistinguishable. The need for symmetry is to calculate the average delay without exactly calculating the queue length distributions. Sidi and Segall in [10] were able to find the average delay of two user symmetric S-ALOHA over the collision channel. They also found the optimal transmission probability to minimize the delay. In [6] authors were able to calculate the average delay of symmetric S-ALOHA over a class of channels with MPR capability, namely, channels with capture. We follow a similar approach to these works to calculate the average delay of S-ALOHA with transmission control without MPR capability. Our results show that if the channel tends to be in 'bad' state more than the 'good' state, then the optimal transmission probability is equal to one over all possible arrival rates, which attracts the attention that not only S-ALOHA with transmission control in this case eliminates the need of scheduling and outperforms TDMA, but also the strategy of transmitting whenever backlogged if the channel is in the 'good' state is both throughput and delay optimal. On the other hand, if the channel tends to be in the 'good' state, then transmission probability equal to one is delay optimal only over a certain range of the arrival rates. We make this more specific in the following lemma:\\
\textit{Lemma 3:}\\
For symmetric S-ALOHA with transmission control under the above assumptions, the average delay is given by:
\begin{equation}
D_{avg}=\frac{(1-\lambda)-\left(1-\frac{\lambda}{2}\right)\pi_1^{(1)}q_{11}}{\pi_1^{(1)}q_{11}f_{11}\left(1-\pi_1^{(1)}q_{11}\right)-\lambda}
\end{equation}
Moreover, the optimal transmission probability that minimizes the delay is given by:\\
\underline{\textit{If $\pi_0^{(1)} \geq 0.5$}}
\begin{equation}
q_{11}^*=1 \mbox{, for  }\lambda \in \left[0,\pi_1^{(1)}\left(1-\pi_1^{(1)}\right)f_{11}\right)
\end{equation}
\underline{\textit{If $\pi_0^{(1)} \leq 0.5$}}
\begin{equation}
q_{11}^* =
\begin{cases}
1 & \textnormal{for  } \lambda \in \left[0, \tilde{\lambda_1}\right),
\\
p_1 & \textnormal{for  } \lambda \in \left[\tilde{\lambda_1},\frac{f_{11}}{4}\right),
\end{cases}
\end{equation}
Where:
 \begin{multline}
 \tilde{\lambda_1}=1\!+\!f_{11}\!\!\left(\!\!1\!-\!2\pi_1^{(1)}\!+\!0.5\left(\pi_1^{(1)}\right)^2\right)-\\
 \sqrt{1\!\!-\!\!\left(\pi_1^{(1)}\right)^2\!\!f_{11}\!\!+\!\!f_{11}^2\!\!\left(\!\!1-\!\!2\pi_1^{(1)}\!\!+\!\!0.5\left(\pi_1^{(1)}\right)^2\!\!\right)^2}
 \end{multline}
\begin{multline}
p_1=\frac{1}{(2-\lambda)\pi_1^{(1)}}\times\bigg\{2(1-\lambda)-\\
\left.\sqrt{4(1\!-\!\lambda)^2\!\!-\!\!\frac{4}{f_{11}}\!\!\left(1-\frac{\lambda}{2}\right)\!\!\left[(1-\lambda)f_{11}\!\!-\!\!\lambda\!\left(1-\frac{\lambda}{2}\right)\right]}\right\}
\end{multline}
\textit{Proof:} See Appendix B.\\
The next two figures compare between the minimum average delay of a symmetric S-ALOHA with and without transmission control for different values of success probabilities $f_{11}$ and different values of stationary probabilities of the channel $\pi_1^{(1)}$ that we denote by p11. The success probability $f_{11}$ only affects the maximum stable arrival rate that can be handled at the queues. On the other hand, the stationary probability plays a major role in the relative advantage of transmission control from a delay point of view: Transmission control has more significant advantage whenever the channel tends to be in the bad state for a longer proportion of time as can be inferred from Lemma 1 and 3.
\begin{figure}[h]
\centering
\includegraphics[width=0.5\textwidth]{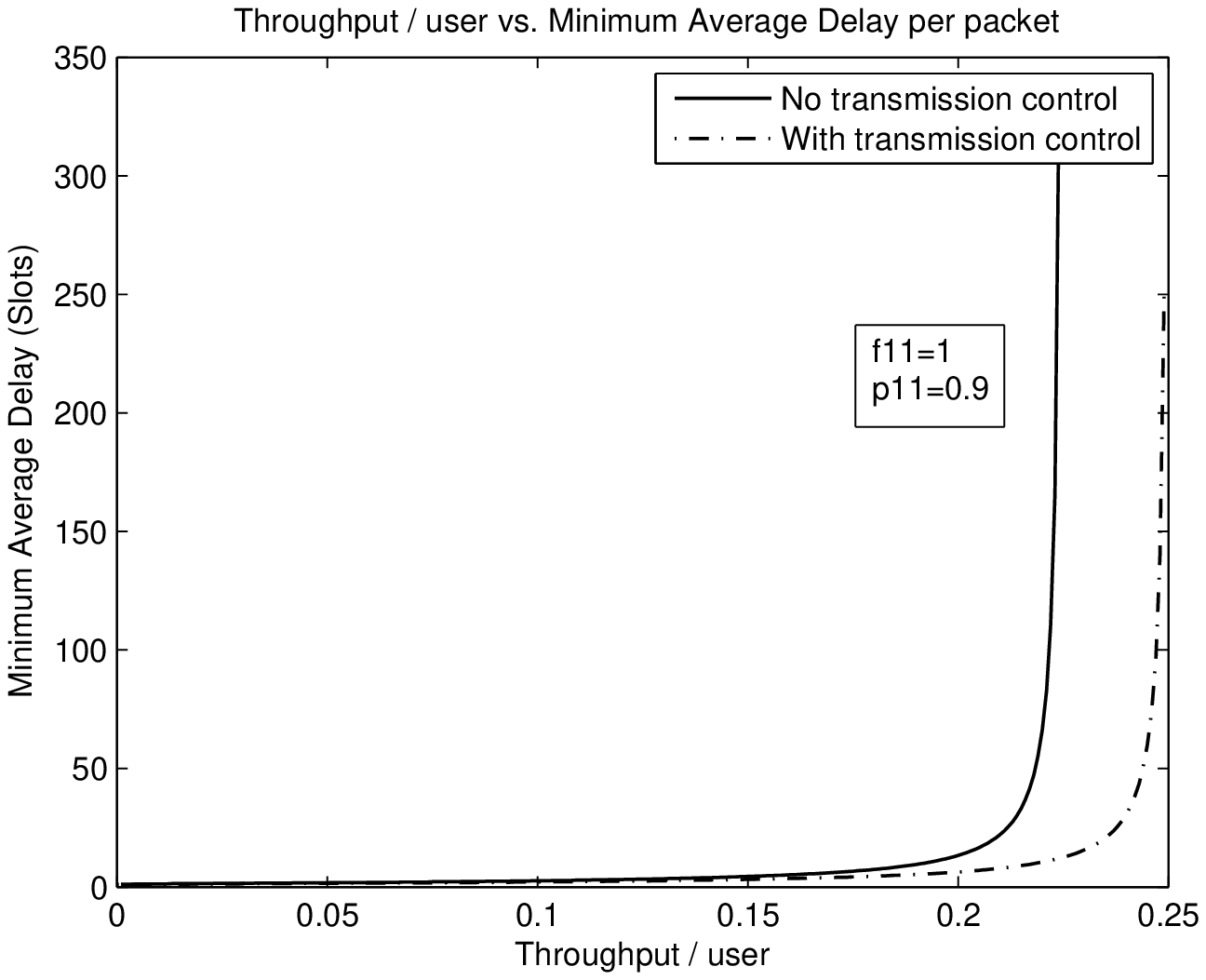}
\vspace{-0.35 in}
\end{figure}

\begin{figure}[h]
\centering
\includegraphics[width=0.5\textwidth]{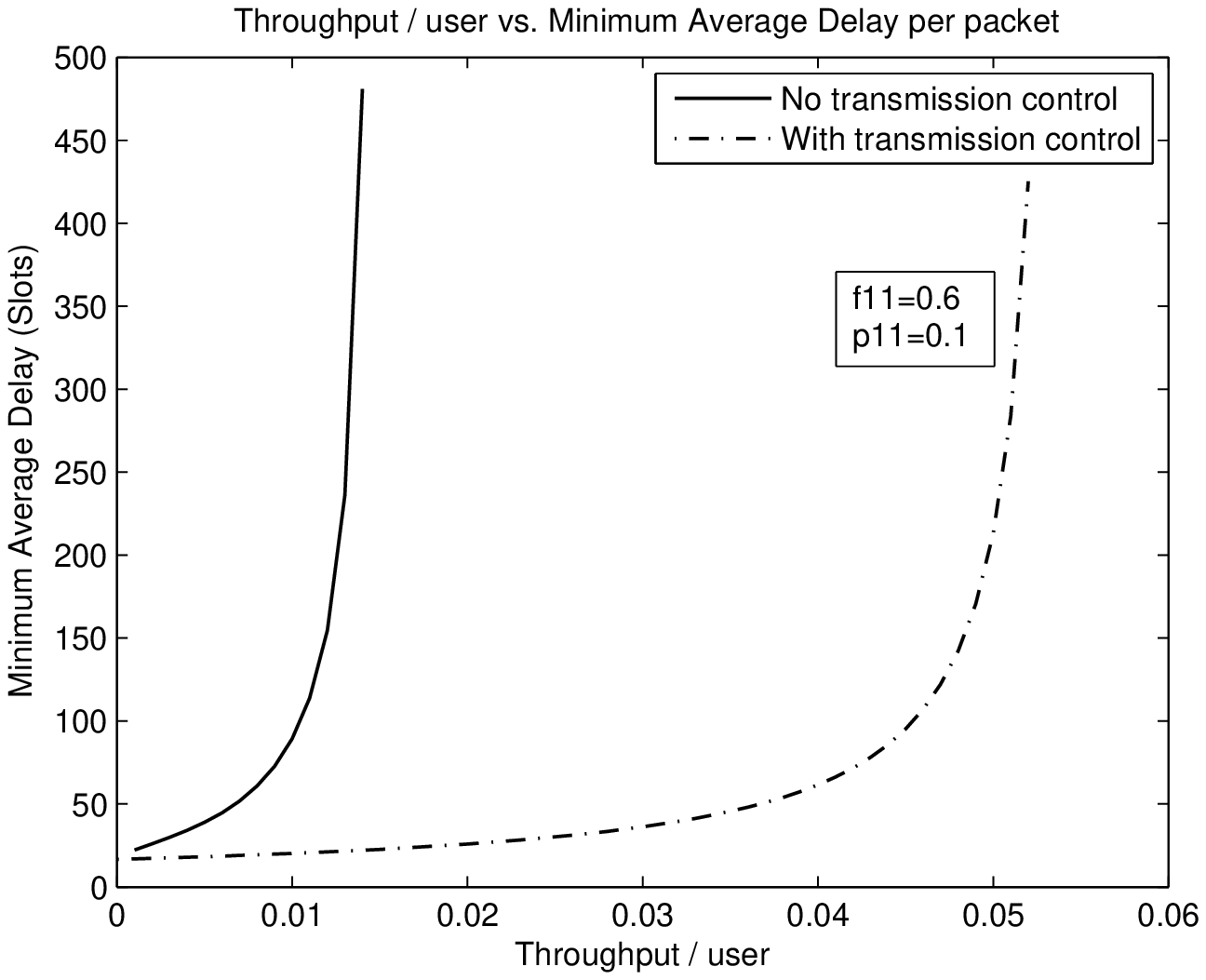}
\vspace{-0.35 in}
\end{figure}

\section{Conclusion}
In this paper, we calculated the stability region and delay for two user random access over Gilbert-Elliott channel in which the users use their knowledge about the channel state to adjust their transmission probabilities. Our results show that random access with transmission control is very effective and outperforms TDMA whenever the channels tend to be in the bad state and the optimal transmission probabilities are one, which eliminates the need of scheduling and hence simplifying the design of MAC-Layer protocol. Moreover, it is shown that these transmission probabilities are delay optimal. If the channels tend to be in the good state, transmission control strictly improves the stability region compared to ordinary S-ALOHA but TDMA is better in this case. Furthermore, enhancing the Physical layer by allowing MPR capability can alleviate this downside, attracting the attention that transmission control can make S-ALOHA very suitable to be implemented over time varying channels in networks lacking the capability of strong coordination between the nodes.

\appendices
\section{Proof of Lemma 1}
In section III, we already found the stability region for a fixed probability pair $(q_{11},q_{12})$ by using the dominant system approach. We use the constrained optimization technique as in [9] to derive the boundary of the stability region. After replacing $\lambda_1$ by $x$ and $\lambda_2$ by $y$, the boundary of the stability region for fixed transmission probability pair can be written as:\\
\begin{eqnarray}
y=\left(\pi_1^{(2)}q_{12}f_{12}\right)\left[1-\frac{x}{f_{11}\left(1-\pi_1^{(2)}q_{12}\right)}\right]\\
\mbox{ for      }0\leq x<\pi_1^{(1)}q_{11}f_{11}\left(1-\pi_1^{(2)}q_{12}\right)
\end{eqnarray}
\begin{eqnarray}
x=\left(\pi_1^{(1)}q_{11}f_{11}\right)\left[1-\frac{y}{f_{12}\left(1-\pi_1^{(1)}q_{11}\right)}\right]\\
\mbox{ for      }0\leq y<\pi_1^{(2)}q_{12}f_{12}\left(1-\pi_1^{(1)}q_{11}\right)
\end{eqnarray}
First we consider the constrained optimization problem as given by equations (31), (32). It can be written as:
\begin{equation}
\max_{q_{12}\in[0,1]}y=\pi_1^{(2)}q_{12}f_{12}-\frac{\pi_1^{(2)}q_{12}f_{12}x}{\left(1-\pi_1^{(2)}q_{12}\right)f_{11}}
\end{equation}
subject to the constraint given by (32).\\
Differentiating with respect to $q_{12}$, we get:
\begin{equation}
\frac{dy}{dq_{12}}=\pi_1^{(2)}f_{12}-\frac{\pi_1^{(2)}f_{12}x}{\left(1-\pi_1^{(2)}q_{12}\right)^2f_{11}}
\end{equation}
Setting (36) to zero, we get:
\begin{equation}
q_{12}^*=\frac{1}{\pi_1^{(2)}}\left(1-\sqrt{\frac{x}{f_{11}}}\right)
\end{equation}
For $q_{12}^*$ to be a valid probability, we should have:
\begin{equation}
\left(1-\pi_1^{(2)}\right)^2f_{11}\leq x \leq f_{11}
\end{equation}
Also, for the constraint in (32) to be satisfied, $x$ must satisfy:
\begin{equation}
x \leq (\pi_1^{(1)})^2f_{11}
\end{equation}
Combining the two conditions, $x$ must satisfy:
\begin{equation}
\left(1-\pi_1^{(2)}\right)^2f_{11}\leq x \leq (\pi_1^{(1)})^2f_{11}
\end{equation}
Substituting in (31), we find that the boundary of the stability region within this range is given by:
\begin{equation}
\sqrt{\frac{\lambda_1}{f_{11}}}+\sqrt{\frac{\lambda_2}{f_{12}}}=1
\end{equation}
Now, we consider the values of $x$ for which $x<\left(1-\pi_1^{(2)}\right)^2f_{11}$. It can be easily shown that:
\begin{equation}
\frac{dy}{dq_{12}}>0, \forall q_{12}\in [0,1].
\end{equation}
Therefore, $q_{12}^*=1$.
For (32) to be satisfied, $x<\pi_1^{(1)}\pi_0^{(2)}f_{11}$.\\
Hence for $0<x<\min\left(\pi_1^{(1)}\pi_0^{(2)}f_{11},\left(1-\pi_1^{(2)}\right)^2f_{11}\right)$
we get by substituting in (31):
\begin{equation}
\frac{\lambda_1}{\pi_0^{(2)}f_{11}}+\frac{\lambda_2}{\pi_1^{(2)}f_{12}}=1
\end{equation}
Finally, for $x>\left(\pi_1^{(1)}\right)^2f_{11}$, and noting that $x<\pi_1^{(1)}f_{11}$, for (32) to be satisfied, we should have that $q_{12}<\frac{1}{\pi_1^{(2)}}\left(1-\frac{x}{\pi_1^{(1)}f_{11}}\right)$. It can be easily shown that over the range $\left(\pi_1^{(1)}\right)^2f_{11}<x<\pi_1^{(1)}f_{11}$, $\frac{dy}{dq_{12}}>0$. Hence, $q_{12}^*=\frac{1}{\pi_1^{(2)}}\left(1-\frac{x}{\pi_1^{(1)}f_{11}}\right)$. For $q_{12}^*$ to be a valid probability, we should have that $\pi_1^{(1)}f_{11}(1-\pi_1^{(2)})<x<\pi_1^{(1)}f_{11}$. Combining both conditions, we get that it is valid for $\max\left(\pi_1^{(1)}(1-\pi_1^{(2)})f_{11},\left(\pi_1^{(1)}\right)^2f_{11}\right)<x<\pi_1^{(1)}f_{11}$.
Substituting in the objective function in (31), we get that for $\max\left(\pi_1^{(1)}(1-\pi_1^{(2)})f_{11},\left(\pi_1^{(1)}\right)^2f_{11}\right)<x<\pi_1^{(1)}f_{11}$:
\begin{equation}
\frac{\lambda_2}{\pi_0^{(1)}f_{12}}+\frac{\lambda_1}{\pi_1^{(1)}f_{11}}=1
\end{equation}
By similar arguments, it can be shown that the other dominant system leads to exactly the same stability region, hence the proof is complete.\\
It should be finally noted that the shape of the stability region depends on whether $\pi_0^{(1)}+\pi_0^{(2)}>1$ or not.
If $\pi_0^{(1)}+\pi_0^{(2)}>1 \Leftrightarrow \pi_1^{(1)}\pi_0^{(2)}f_{11} < \left(1-\pi_1^{(2)}\right)^2f_{11} \Leftrightarrow \pi_1^{(1)}(1-\pi_1^{(2)})f_{11}>\left(\pi_1^{(1)}\right)^2f_{11}$,
the stability region consists of two linear parts while if
$\pi_0^{(1)}+\pi_0^{(2)}<1 \Leftrightarrow \pi_1^{(1)}\pi_0^{(2)}f_{11} > \left(1-\pi_1^{(2)}\right)^2f_{11} \Leftrightarrow \pi_1^{(1)}(1-\pi_1^{(2)})f_{11}<\left(\pi_1^{(1)}\right)^2f_{11}$,
the stability region consists of three parts as in Lemma 1.
\section{Proof of Lemma 3}
Our proof follows a similar approach to [6] and [10] in order to solve for the average delay in a symmetric configuration without explicitly solving for the joint queue statistics.
The queues of both users evolve as:
\begin{equation}
Q_i^{t+1}=\left[Q_i^t-D_i^t\right]^++A_i^t, \mbox{  } i=1,2
\end{equation}
Where $Q_i^t$ is the queue length of user $i$ at any time slot $(t)$, $A_i^t$ is the number of arrivals to user $i$ queue at time slot $(t)$ and $D_i^t$ is the number of departures from user $i$ queue at time slot $(t)$.\\
To calculate the delay, we solve for the moment generating function of the joint queue lengths of $Q_1$ and $Q_2$ denoted by $G(x,y)=\lim_{t \to \infty}\textbf{E}\left[x^{Q_1^t}y^{Q_2^t}\right]$. By using the queue evolution equation and denoting by $\lambda$ the average arrival rate to each queue in the symmetric configuration, we get:
\begin{multline}
\textbf{E}\left[x^{Q_1^t}y^{Q_2^t}\right]=\textbf{E}\left[x^{A_1^t}y^{A_2^t}\right].\bigg\{\textbf{E}\left[\textbf{1}\left[Q_1^t=0,Q_2^t=0\right]\right]+\\
\textbf{E}\left[x^{Q_1^t-D_1^t}\textbf{1}\left[Q_1^t>0,Q_2^t=0\right]\right]+\\
\textbf{E}\left[y^{Q_2^t-D_2^t}\textbf{1}\left[Q_1^t=0,Q_2^t>0\right]\right]+\\
\textbf{E}\left[x^{Q_1^t-D_1^t}y^{Q_2^t-D_2^t}\textbf{1}\left[Q_1^t>0,Q_2^t>0\right]\right]\bigg\}
\end{multline}
Taking the limit as $t \to \infty$, we get:
\begin{multline}
G(x,y)=F(x,y).\bigg\{G(0,0)+\\
\left[G(x,0)-G(0,0)\right].\left[\frac{\pi_1^{(1)}q_{11}f_{11}}{x}+\left(1-\pi_1^{(1)}q_{11}f_{11}\right)\right]+\\
\left[G(0,y)-G(0,0)\right].\left[\frac{\pi_1^{(1)}q_{11}f_{11}}{y}+\left(1-\pi_1^{(1)}q_{11}f_{11}\right)\right]+\\
\left[G(x,y)-G(x,0)-G(0,y)+G(0,0)\right].\\
\Big\{\left[\left(\pi_1^{(1)}q_{11}f_{11}\right)\left(1-\pi_1^{(1)}q_{11}\right)\left(\frac{1}{x}+\frac{1}{y}\right)\right]+\\
\left[1-2\left(\pi_1^{(1)}q_{11}f_{11}\right)\left(1-\pi_1^{(1)}q_{11}\right)\right]\Big\}\bigg\}
\end{multline}
Where:
\begin{equation}
F(x,y)=(\lambda x+1-\lambda)(\lambda y+1-\lambda)
\end{equation}
Using that $G(1,1)=1$ and $G(0,1)=G(1,0)$ by symmetry, and using L'H\^{o}pital rule, we get:
\begin{multline}
G(0,0)\!\left[\left(\pi_1^{(1)}\right)^2\!q_{11}^2f_{11}\right]\!+G(1,0)\!\left[\pi_1^{(1)}q_{11}f_{11}\right]\!\!\!\left[1\!-\!2\pi_1^{(1)}\!q_{11}\right]\\
=\left[\pi_1^{(1)}q_{11}f_{11}\left(1-\pi_1^{(1)}q_{11}\right)-\lambda\right]
\end{multline}
Calculating $G_1(1,1)=\frac{\partial G(x,y)}{\partial x}$ at $(x,y)=(1,1)$, we get by using L'H\^{o}pital rule:
\begin{multline}
\left[\pi_1^{(1)}q_{11}f_{11}\left(1-\pi_1^{(1)}q_{11}\right)\right]G_1(1,1)=\\
\lambda(1-\lambda)-G_1(1,0)\left(\pi_1^{(1)}\right)^2\!q_{11}^2f_{11}
\end{multline}
Calculating $\frac{dG(x,x)}{dx}$ at $x=1$ and using L'H\^{o}pital rule, we get:
\begin{multline}
\frac{d}{dx}G(x,x)\bigg|_{x=1}=\frac{G_1(1,0)\left(\pi_1^{(1)}q_{11}f_{11}\right)\left(1-2\pi_1^{(1)}q_{11}\right)}{\left[\pi_1^{(1)}q_{11}f_{11}\left(1-\pi_1^{(1)}q_{11}\right)-\lambda\right]}+\\
\frac{2\lambda-3\lambda^2}{2\left[\pi_1^{(1)}q_{11}f_{11}\left(1-\pi_1^{(1)}q_{11}\right)-\lambda\right]}
\end{multline}
Using that $\frac{dG(x,x)}{dx}\bigg|_{x=1}=2G_1(1,1)$ and after some manipulations, we get:
\begin{equation}
G_1(1,1)=\frac{\lambda(1-\lambda)+\pi_1^{(1)}q_{11}\left(\frac{\lambda^2}{2}-\lambda\right)}{\left[\pi_1^{(1)}q_{11}f_{11}\left(1-\pi_1^{(1)}q_{11}\right)-\lambda\right]}
\end{equation}
By using Little's law, we get the average delay per packet as:
\begin{equation}
D_{avg}=\frac{G_1(1,1)}{\lambda}=\frac{(1-\lambda)+\pi_1^{(1)}q_{11}\left(\frac{\lambda}{2}-1\right)}{\left[\pi_1^{(1)}q_{11}f_{11}\left(1-\pi_1^{(1)}q_{11}\right)-\lambda\right]}
\end{equation}
We next seek $q_{11}$ which minimizes $D_{avg}$ while conserving stability. Specifically, we need to solve:
\begin{multline}
\min_{q_{11}\in[0,1]}D_{avg}=\frac{(1-\lambda)+\pi_1^{(1)}q_{11}\left(\frac{\lambda}{2}-1\right)}{\left[\pi_1^{(1)}q_{11}f_{11}\left(1-\pi_1^{(1)}q_{11}\right)-\lambda\right]}\\
\mbox{s.t.     }\lambda<\pi_1^{(1)}q_{11}f_{11}\left(1-\pi_1^{(1)}q_{11}\right)\qquad \quad
\end{multline}
Consider the constraint for stability, the constraint can be written as:
\begin{equation}
\left(\pi_1^{(1)}\right)^2q_{11}^2f_{11}-\pi_1^{(1)}f_{11}q_{11}+\lambda<0
\end{equation}
The roots of this equation that we denote by $s_1$ and $s_2$ are given by:
\begin{equation}
s_1,s_2=\frac{1\mp\sqrt{1-4\lambda/f_{11}}}{2\pi_1^{(1)}}
\end{equation}
Hence, the stability constraint implies that the optimal probability $q_{11}^*$ satisfies $s_1 \leq q_{11}^* \leq s_2$.
Ignoring for the moment the constraints and equating the derivative of the objective function to zero, we get that the optimal transmission probabilities are given by:
\begin{multline}
p_1,p_2=\frac{(1-\lambda)\mp\sqrt{\lambda/2}\sqrt{\frac{2}{f_{11}}\left(1-\lambda/2\right)^2-(1-\lambda)}}{\left(1-\lambda/2\right)\pi_1^{(1)}}
\end{multline}
After some algebraic manipulations, we can show that $0 \leq s_1 \leq 1$ and that $s_1 \leq p_1 \leq s_2 \leq p_2$. As the objective function is strictly decreasing on $(s_1,p_1)$, we can conclude that the optimal transmission probability $q_{11}^*$ to minimize the delay is given by: $q_{11}^*=\min(p_1,1)$.\\
The stability condition yields that $\lambda<\lambda_{\max}$, where:
\begin{equation}
\lambda_{\max} =
\begin{cases}
f_{11}/4 & \textnormal{if  } \pi_1^{(1)} \geq 1/2 \Leftrightarrow \pi_0^{(1)} \leq 1/2,
\\
\pi_1^{(1)}\left(1-\pi_1^{(1)}\right)f_{11} & \textnormal{if  } \pi_1^{(1)} \geq 1/2 \Leftrightarrow \pi_0^{(1)} \leq 1/2,
\end{cases}
\end{equation}
It can be shown after some manipulations that $p_1>1 \Leftrightarrow \lambda \in \left[0,\lambda^*\right]$ and that $p_1 < 1 \Leftrightarrow \lambda \in \left[\lambda^*,\lambda_{\max}\right)$, where:
\begin{multline}
\lambda^*=1+f_{11}\left(1-2\pi_1^{(1)}+\left(\pi_1^{(1)}\right)^2/2\right)-\\
\sqrt{1-\left(\pi_1^{(1)}\right)^2f_{11}+f_{11}^2\left(1-2\pi_1^{(1)}+\left(\pi_1^{(1)}\right)^2/2\right)^2}
\end{multline}
Noting that $\lambda^*(\pi_1^{(1)}=1/2)=f_{11}/4$ and that $\lambda^*<\lambda_{\max}=f_{11}/4$ only if $\pi_0^{(1)} \leq 1/2$, the proof is complete.

    \end{document}